\documentclass[aps,twocolumn,showpacs,floatfix,prb]{revtex4}
\usepackage{graphicx}
\usepackage{amsmath}
\begin{document}
\title{Dispersion Anomalies in Cuprate Superconductors}
\author{A. V. Chubukov$^1$ and M. R. Norman$^2$}
\affiliation{$^1$ Department of Physics, University of Wisconsin,
Madison, WI 53706} 
\affiliation{$^2$Materials Science Division, Argonne National Laboratory, Argonne,
IL 60439}
\date{\today}
\begin{abstract}
We argue that the shape of the dispersion
  along the nodal and  antinodal directions in the cuprates
  can  be understood as a consequence of the
   interaction of the electrons with collective spin excitations.
 In the normal state, the dispersion
 displays a crossover at an energy  where the decay into spin fluctuations 
becomes relevant. In the superconducting state, the antinodal dispersion 
 is strongly affected by the $(\pi,\pi)$ spin resonance and displays
 an $S-$shape whose magnitude scales with the resonance intensity. 
For nodal fermions, relevant spin  excitations do not have resonance
 behavior, rather they are better characterized as a gapped continuum.
 As a consequence, the $S-$shape becomes a kink,
 and superconductivity does not affect the dispersion as strongly.
Finally, we note that optical phonons typically lead to a temperature
independent $S-$shape, in disagreement with the observed dispersion.
\end{abstract}
\pacs{74.25.Jb, 74.72.Hs, 79.60.Bm, 74.50.+r}

\maketitle

\section{Introduction}

Angle resolved photoemission (ARPES) experiments are valuable sources
 of information about the shape of the Fermi surface in the cuprates,
 and the frequency, momentum and temperature dependence of the 
electron self-energy.
The subject of this paper is an analysis of the
 dispersion along various momentum cuts. 
 These dispersions have been obtained by several 
 groups \cite{valla,bogdanov,kaminski,lanzara,johnson,ronning,zhou,sato,kim,gromko} by
 high precision  measurements of the momentum distribution curves (MDCs), by which the 
 spectral function 
 is obtained at a given energy by
 making  scans along directions normal to the Fermi surface.  
The spectral function $A({\bf k}, \omega) = (1/\pi) |Im G({\bf k}, \omega)|$ is
 related to the self-energy $\Sigma_{\bf k} (\omega)$ as
\begin{equation}
A({\bf k}, \omega) = \frac{1}{\pi} \frac{\Sigma_{\bf k}^{\prime \prime}
 (\omega)}{(\omega - \epsilon_{\bf k} -
\Sigma_{\bf k}^{\prime} (\omega))^2 + (\Sigma_{\bf k}^{\prime \prime} (\omega))^2}
\label{1}
\end{equation}
Near the Fermi surface,  
$\epsilon_{\bf k} \approx v_F ({\bf k}_F) (k_\perp - k_F)$ where $v_F ({\bf k}_F)$
 is the bare value of the velocity. 
There are several reasons (both theoretical and experimental) to believe
 that the self-energy weakly depends on 
 the value of $k_{\perp}$  normal to 
the Fermi surface, and can be approximated as $\Sigma_{\bf k} (\omega) \approx
 \Sigma_{{\bf k}_F} (\omega)$. For a given ${\bf k}_F$ 
specified by a cut, the 
 MDC spectral function $A (k_\perp, \omega = const)$ 
   is then 
a Lorentzian centered at $v_F ({\bf k_F}) (k_\perp - k_F) = \omega - 
\Sigma_{{\bf k}_F} (\omega)$ with a HWHM equal to  
$\Sigma^{\prime \prime}_{{\bf k}_F} (\omega)/v_F ({\bf k_F})$ 
(Ref.~\onlinecite{kaminski}). 
In a generic Fermi liquid, the self-energy is linear in $\omega$  at the lowest 
energies: $\Sigma^{\prime}_{{\bf k}_F} (\omega) = -\lambda_{{\bf k}_F} \omega$. 
The position of the MDC peak then determines
 the renormalized Fermi velocity $v^*_F = v_F/(1 + \lambda_{{\bf k}_F})$. 
At higher energies, $\lambda$ becomes frequency dependent, and the 
dispersion deviates from the linear form.

The MDC data have revealed several  
characteristic  features of the dispersion 
which  need to be  explained:
\begin{itemize}
\item In the normal state, the dispersion along both the nodal and antinodal 
 directions 
shows a relatively smooth 
crossover from a linear behavior at small binding energies to a
 more steep behavior above roughly $50-70 meV$. 
This effect has been observed in $Bi2212$ \cite{valla,bogdanov,kaminski,lanzara,johnson,sato},
 $Bi2201$ \cite{lanzara,sato}, 
$Bi2223$ \cite{sato}, $Na_xCCOC$ \cite{ronning}, and $LSCO$ \cite{lanzara,zhou}. In the last case,
 the crossover is sharper and more resembles a kink.   
 \item  The renormalized 
Fermi velocity $v^*_F$  along the nodal direction  
  weakly depends on doping \cite{zhou} and in $Bi2212$ 
equals $1.6 eV\AA$ \cite{kaminski,zhou}. At the same time, 
 at high energies, the dispersion is strongly
doping dependent, becoming steeper with 
underdoping \cite{johnson,zhou}.
\item  In the superconducting state, the dispersion along the antinodal direction 
 develops an $S-$shape \cite{mike01}, with a ``negative'' dispersion 
 between $60$ and $80 meV$.  This feature develops with decreasing temperature in an order parameter like fashion,
 with an onset temperature at $T_c$ for overdoped samples, and somewhat above $T_c$ for underdoped
 samples \cite{sato,gromko}.
\item The dispersion along the nodal 
direction does not 
develop an $S-$shape below $T_c$. Instead, in the superconducting state 
 the crossover gets sharper, with a
 kink-like feature near $70 meV$ developing
 in $Bi2212$ \cite{bogdanov,kaminski,lanzara,johnson,sato} and $Bi2223$ \cite{sato}.
This extra ``sharpness" has a temperature dependence similar to that of the antinodal
dispersion mentioned above \cite{johnson}.
In $Bi2201$ and $LSCO$, on the other hand, the nodal 
dispersion does not change much between the normal and superconducting states \cite{lanzara,sato}.
\item The high energy nodal dispersion never recovers to the bare dispersion.  It remains linear to the highest binding
energy studied \cite{ronning}, with an interpolation to a ${\bf k}$ point at zero energy displaced well inside the
Fermi surface.
\end{itemize}    

Theoretical scenarios proposed to explain the data 
 differ primarily on  whether the electron-electron 
or the electron-phonon interaction 
 is responsible for the observed behavior.  
In the electron-electron scenario, the crossover from
a linear dispersion at the lowest energies
to a more steep dispersion at higher energies
has been identified \cite{acs} with  the crossover
 from Fermi liquid 
 to non-Fermi liquid behavior.
In the superconducting state, the $S-$shape
 dispersion along the antinodal direction has been 
associated \cite{ding,ac,eschrig,mike01} with the interaction 
with the $(\pi,\pi)$ spin exciton, 
 which  below $T_c$ emerges at a frequency $\omega_{res} < 2\Delta$  due 
to a feedback of the pairing on the spin susceptibility.  
The interaction with the exciton gives rise to a $\Sigma^\prime (\omega)$ which
 strongly increases and then rapidly drops
 as $\omega$ approaches  $\omega_0 = \Delta + \omega_{res}$. This gives 
rise to the observed $S-$shape dispersion.  
In optimally doped $Bi2212$, both $\Delta$ and 
$\omega_{res}$ are close to $40 meV$ \cite{pdh},
 i.e., $\omega_0 \sim 80 meV$. The 
 magnitude of the $S-$shape piece  is stronger  for $|\omega| < \omega_0$, since for $|\omega| >\omega_0$, 
 $\Sigma^{\prime \prime}$ rapidly increases.  Both the  value of $\omega_0$ 
 and the asymmetry of the $S-$shape dispersion   agree
 with the data \cite{mike01,sato,gromko}. 

For nodal fermions, scattering by ${\bf q_0} = (\pi,\pi)$  shifts the nodal Fermi point
to an energy about $0.7 eV$ above $E_F$. This energy is 
 too high to expect any appreciable effect on the low energy dispersion.   
Still, Eschrig and Norman \cite{eschrig}
 argued that the kink near $70 meV$ can be explained by the interaction 
 with the spin resonance, as the resonance has a finite momentum 
width around $(\pi,\pi)$ \cite{keimer}. 
 They used a phenomenological form of the 
 spin susceptibility 
 with a sharp $\delta$ function in frequency at $\omega_{res}$ and 
 a Lorentzian in momentum space with a width of 2 lattice constants,
  with the momentum smearing giving rise to resonance scattering
 between the node and other Fermi surface points. 
 But a good description of the nodal fermion spectrum required the inclusion
 of a gapped continuum in the spin excitation spectrum, which acts to smear
 the $S-$shape into a kink, and also gives rise to the linear $\omega$ behavior in $Im\Sigma$
 observed at higher binding energies.

 The electron-phonon scenario for the dispersion was put forward in Ref.~\onlinecite{lanzara}.
 The key difference with the electron-electron scenario is in the interpretation of the normal state ARPES data --
 Ref.~\onlinecite{lanzara} argued that there is a sharp kink (rather than a crossover) in the dispersion 
 in the normal state, and that the  kink energy is 
 about the same in all materials studied ($LSCO$, $Bi2201$, and $Bi2212$). 
They further argued that the kink effect is rather isotropic in the Brillouin zone (in disagreement
with other work \cite{kaminski,sato,gromko}). 
They speculated that this similarity implies that 
superconductivity plays a secondary  role in the phenomenon, and that 
the features in the dispersion can be reproduced 
 by coupling an electron to a  bosonic mode unrelated to 
superconductivity. They suggested a $(\pi,0)$  optical phonon 
with an energy of 
$55 meV$ \cite{mcqueeny} as the best candidate.

In this paper, we distinguish between these two possibilities and argue in favor of a
spin-fluctuation  scenario. 
We  first analyze  the spin-fluctuation  scenario in  more detail. 
We argue that the spin resonance scattering is effective in scattering 
antinodal fermions  near the Fermi energy, but is not effective for nodal fermions
since the bosonic momenta which connect a nodal point with  other points on the Fermi surface
are far removed from $(\pi,\pi)$.
 Rather, the most effective low-energy scattering for a nodal fermion is to the antinode, 
 where the density of states has a singularity. 
This scattering  gives rise to a kink in the self-energy of a nodal fermion   
 at $\sim 2\Delta$ which
 generates a kink in the dispersion at the same energy. We 
 find that the resonance scattering emerges away from the nodal direction, and 
  the magnitude of the resulting $S-$shape
 dispersion  progressively increases as the antinode is approached \cite{gromko}.
 
 We then argue that the interaction 
 with an Einstein phonon  gives rise to a  temperature independent $S-$shape 
 dispersion for all cuts normal to the Fermi surface.
This is difficult to reconcile with both the antinodal dispersion, for which the $S-$shape 
is present but only emerges below $T_c$, and the
 nodal dispersion which does not display an  $S-$shape form 
 at any temperature.   

\section{ Magnetic scattering} 

In the magnetic scenario, the fermionic self-energy 
 originates from the strong spin component of the
electron-electron interaction in the particle-hole channel 
 and 
 can be viewed as coming from scattering by collective spin fluctuations.
 To lowest order,
 the corresponding self-energy is given by  
\begin{equation}
\Sigma_{{\bf k_F}} (\omega) =  -\frac{3i g_{s}^2}{8 \pi^3}~
 \int d^2 q d \Omega G^0_{{\bf k}_F+{\bf q}} 
 (\omega + \Omega) \chi_{s} ({\bf q}, \Omega)   
\label{s1}
\end{equation}
where $g_s$ is the spin-fermion coupling.
Here $G^0$ is the bare Green's function
 (in the normal state, $G^{0}({\bf k}, \omega) = 
1/(\omega - v_F ({\bf k_F}) (k_{\perp} - k_F))$), 
and $\chi_s ({\bf q}, \omega)$ is the dynamical spin susceptibility
for which one has, 
\begin{equation}
\chi^{-1}_s ({\bf q}, \Omega) = \chi^{-1}_s ({\bf q}) - \Pi_{\bf q} (\Omega)
\end{equation}
where $\chi_s ({\bf q})$ is  the static part of the susceptibility which
 is believed to be peaked at or near the antiferromagnetic momentum ${\bf q_0} = (\pi,\pi)$,
  and $\Pi_{\bf q} (\Omega)$ (subject to $\Pi_{\bf q} (0) =0$) accounts for the spin dynamics
and is proportional to the dynamical part of the full particle-hole 
 bubble. 
For a non-diagonal ${\bf q}$ that connects points on the Fermi surface,
 the polarization operator has the 
form \cite{acs2}
\begin{equation}
\Pi_{\bf q} (\Omega) = i \sum_m \frac{\gamma_{\bf q}}{2}~\int_{-\infty}^{\infty} d \omega 
\left[1-\frac{\Delta_+ \Delta_- + \omega_+\omega_-}{\sqrt{\omega^2_+ 
-\Delta^2_+} \sqrt{\omega^2_- -\Delta^2_-}}\right].
\label{s2}
\end{equation}
Here $\omega_\pm = \omega \pm \Omega/2$ and 
$\Delta_\pm = \Delta ({\bf k}_m \pm {\bf q}/2)$
 are the values of the  $d-$wave gap at the points ${\bf k}_m \pm {\bf q}/2$
 which are simultaneously on the Fermi surface,  
the summation $m$ being over a discrete set of these points. The prefactor  
 $\gamma_{\bf q}$ depends on the coupling $g_s$ 
 and the angle between the Fermi velocities at ${\bf k}_m \pm {\bf q}/2$ \cite{acs}.
In principle, the pairing gap $\Delta$ depends on frequency, but
 this dependence is not essential for our purposes and  
we neglect it  for clarity. 

\subsection{Normal state}

\subsubsection{polarization operator}

In the normal state $\Delta =0$, and $\Pi_{\bf q} (\Omega)$ is purely imaginary:
$\Pi_{\bf q} (\Omega) = i \gamma_{\bf q} |\Omega|$. This is the expected result as
once ${\bf q}$ is such that two ${\bf k}-$points separated by ${\bf q}$ can be 
 simultaneously put on the Fermi surface, the polarization bubble contains  
 a Landau damping term. This term generally has the form 
$i|\Omega|/\sqrt{(v_F q)^2 -\Omega^2}$ but in our case $q$ is finite and $v_F q$ 
well exceeds $\Omega$. The true polarization bubble also contains a frequency 
independent piece, but this piece comes from fermions with energies comparable to
 the bandwidth and is already incorporated into $\chi^{-1}_s ({\bf q})$. 
Note that this separation is consistent with 
 a Kramers-Kronig (KK) analysis: a KK transformation of 
$Im \Pi (\Omega) =\gamma  \Omega$ does not  produce a 
 universal piece of $Re \Pi (\Omega)$ independent of the upper limit of the frequency integration.

\subsubsection{fermionic self-energy}

\begin{figure}[tbp]
\includegraphics[width=\columnwidth]{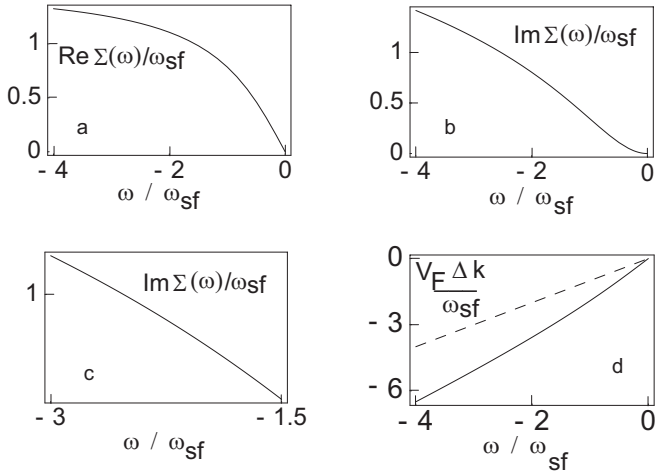}
\caption{ 
The self-energy and  the
dispersion in the normal state for a flat static susceptibility 
$\chi_s ({\bf q})$ near $\bf{q_0} = (\pi,\pi)$. 
a) $Re\Sigma_N$, 
b) and c) $Im \Sigma_N$, with c)
 over intermediate frequencies where  $Im \Sigma_N (\omega)$  displays a
 quasi-linear behavior. 
 d) dispersion $\omega - Re \Sigma (\omega) = v_F \Delta k$ ($\Delta k = k-k_F$), 
 with the dashed line the free fermion dispersion. 
 The coupling is $\lambda =2$.}
\label{fig1a}
\end{figure}

Substituting the relaxational $\chi_s ({\bf q}, \Omega)$ into the self-energy, introducing a 
small ${\bf q}$ via ${\bf q} \rightarrow {\bf q} + {\bf q_0}$  and 
 linearizing the fermionic dispersion near the Fermi surface,
we obtain from (\ref{s1})
\begin{widetext}
 \begin{equation}
\Sigma_{N,{\bf k}_F} (\omega) = -i \frac{3 g_s^2}{8\pi^3}
 \int d q_{\parallel} d q_{\perp} d \Omega~\frac{1}{ \omega+ \Omega - 
v_F({\bf k}_F + {\bf q} + {\bf q_0}) q_{\perp} + i\delta_{\omega + \Omega}}~
 \frac{1}{\chi^{-1}_s ({\bf q} + {\bf q_0}) - i \gamma_{{\bf q} + {\bf q_0}} |\Omega|}
\label{s28} 
\end{equation}
\end{widetext}
where $N$ stands for normal state. For consistency with the assumption that
 the self-energy weakly depends on $\epsilon_{\bf k}$, 
we assume that the fermionic propagator changes much faster with $q_{\perp}$
 than the bosonic $\chi ({\bf q}, \Omega)$, i.e., that the 
 Fermi velocity is much larger than the ``spin'' velocity. We then 
 integrate over momentum $q_{\perp}$ normal to the Fermi surface only 
in the fermionic propagator, and  set  $q_{\perp}$ in the bosonic propagator
 equal to its value at a distance 
between ${\bf k}_F$ and some other point on the Fermi 
surface which is parametrized by  $q_{\parallel}$.
 The integration  over $q_\perp$ is straightforward, and performing it 
 using the fact that $\chi_s ({\bf q}, \Omega)$ is an even function of frequency,
 we  obtain 
 \begin{eqnarray}
&&\Sigma_{N,{\bf k}_F} (\omega) = -
 \frac{3 g_s^2}{4\pi^2}
 \int d q_{\parallel} \frac{1}{v_F ({\bf k}_F + {\bf q_0} + 
{\bf q}_{\parallel})}~\nonumber \\
&&  \int_0^\omega \frac{d \Omega}{\chi^{-1}_s ({\bf q_0}+{\bf q}_{\parallel}) - 
i \gamma_{{\bf q_0}+{\bf q}_\parallel} \Omega}
 \label{s34} 
\end{eqnarray}
The remaining integral over $q_{\parallel}$ 
depends on  the momentum dispersions in $\chi_s ({\bf q})$ 
and  $v_F ({\bf k_F})$ along the Fermi surface, 
both are inputs for the low-energy theory.
For a qualitative understanding of the crossover in the dispersion, 
we assume momentarily that $\chi_s ({\bf q_0}+{\bf q}_\parallel)$ is flat near 
${\bf q_0}$, and that 
 $\gamma_{{\bf q_0}+{\bf q}_\parallel}$ and $v_F ({\bf k}_F + {\bf q_0} + {\bf q}_\parallel)$
 are also momentum independent.
We then immediately obtain from (\ref{s34})  
\begin{equation}
\Sigma_{N, {\bf k}_F} (\omega) = -i\lambda \omega_{sf} \log{\left[1-\frac{i 
|\omega|}{\omega_{sf}}\right]}~ sgn (\omega)
\label{s6}
\end{equation}
where $\lambda = 3 g_s^2 \chi_s/(2\pi v_F)$ and $\omega_{sf} = (\chi_s \gamma)^{-1}$ 
(we use the same notation as in earlier works \cite{pines,acs}). 
Separating real and imaginary parts of the complex logarithm, we obtain
from (\ref{s34})  
\begin{eqnarray}
&& Re \Sigma_{N, {\bf k}_F} (\omega) = -\lambda \omega_{sf} 
\arctan {\frac{\omega}{\omega_{sf}}} \nonumber \\
&&
Im \Sigma_{N, {\bf k}_F} (\omega) = -\lambda \frac{\omega_{sf}}{2}~
 \log{\left[1+\frac{\omega^2}{\omega^2_{sf}}\right]}~ sgn (\omega)
\label{s61}
\end{eqnarray}
At low frequencies, 
 one indeed  recovers Fermi liquid behavior:
\begin{equation}
\Sigma_{N, {\bf k}_F}  (\omega) = - 
\lambda \left(\omega + i \frac{\omega |\omega|}{2\omega_{sf}}\right) 
\label{s343}
\end{equation}
On the other hand, at frequencies larger than $\omega_{sf}$, the self-energy 
 nearly saturates 
\begin{equation}
\Sigma^\prime_N (\omega) \approx - (\pi/2) \lambda \omega_{sf} sgn( \omega),~~~
\Sigma^{\prime \prime}_N (\omega) \propto \log |\omega|
\label{need1}
\end{equation} 
The evolution of $\Sigma^\prime_N (\omega)$ with frequency gives rise to a
 crossover in  the
 normal state dispersion  $\omega - \Sigma_N^\prime (\omega) = v_F (k_\perp -k_F)$
 around $\omega = \omega_{sf}$. We illustrate this in  Fig.\ref{fig1a}.  
We clearly see that the dispersion is linear  below $\omega_{sf}$, with the 
effective velocity $v_F^* = v_F /(1 + \lambda)$. However, above $\omega_{sf}$,
 the dispersion crosses over to a more steep form which also yields an intercept 
 at a finite $k_\perp-k_F$ if extrapolated formally 
to zero energy. This crossover behavior is consistent with
 the one observed experimentally.  Note also that $\Sigma_N^{\prime \prime} (\omega)$ 
 is almost linear in frequency in a relatively wide frequency range above $\omega_{sf}$. 
 This quasi-linearity seems to be a generic property of
 $\Sigma_N^{\prime \prime}$ in the crossover region between $\omega^2$ Fermi liquid 
 behavior at small frequencies  and quantum-critical, non-Fermi-liquid behavior at larger frequencies.
 
\begin{figure}[tbp]
\includegraphics[width=\columnwidth]{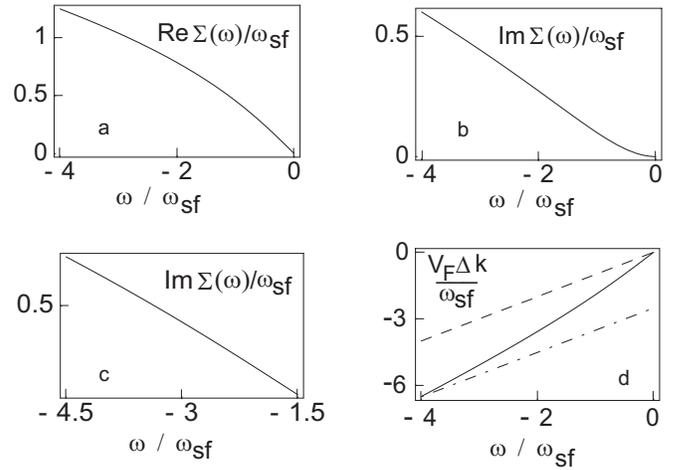}
\caption{Same as Fig.~\protect\ref{fig1a}, but for the Ornstein-Zernike form of the static spin susceptibility. 
In both cases, the crossover occurs around $\omega = \omega_{sf}$
The dashed line in d) is
the free fermion dispersion.
The dashed-dotted line in d) is an extrapolation from high frequencies. 
The extrapolated dispersion crosses the vertical axis at a negative $\Delta k$, i.e., for $k$ inside the Fermi surface.}
\label{fig1b}
\end{figure}

The inclusion of the momentum dependences of
$\chi_s ({\bf q})$, $\gamma_{\bf q}$ and $v_F 
({\bf k}_F)$ gives rise to some $\omega$ dependence of $\Sigma^\prime_N (\omega)$ 
at high  frequencies, and to the angular dependence of the coupling constant $\lambda$, but the
 crossover near $\omega_{sf}$ still survives. 
To illustrate this, in  Fig.\ref{fig1b} we plot the 
 the dispersion obtained for the  
Ornstein-Zernike form of $\chi^{-1}_s ({\bf q}) \propto 
 1 + (q \xi)^2$ with constant $\gamma$ and 
 $v_F$.
We see that the dispersion is again linear at small $\omega$ 
with $v^*_F = v_F/(1+ \lambda_{{\bf k}_F})$, and crosses over to a more steep
 dispersion above $\omega_{sf} ({\bf k}_F)$. Observe  also that 
 $\Sigma^{\prime \prime} (\omega)$ is again nearly linear above $\omega_{sf} ({\bf k}_F)$.
The crossover frequency 
$\omega_{sf} ({\bf k}_F)= (\gamma\chi_s ({\bf k}_F))^{-1}$ is 
 smallest for an antinodal fermion  and largest for a nodal fermion 
simply because the node-node distance is smaller than ${\bf q_0} = (\pi,\pi)$,
 and hence  for node-node scattering 
$\chi^{-1} _s ({\bf q}) \propto (1 + (|{\bf q} - {\bf q_0}|\xi)^2)$.
 For an antinodal fermion, on the
 other hand, the antinode-antinode scattering involves momenta
 very close to ${\bf q_0}$, hence $\chi^{-1}_s  ({\bf q})$ is smaller as the piece 
$(|{\bf q}-{\bf q_0}|\xi)^2$ is absent.
This effect is, however, partly compensated by the fact that $\gamma_{{\bf q}}$
is enhanced around a nodal point and formally diverges for node-node
 scattering because the Landau damping blows up  when the 
velocities of the two fermions in the particle-hole bubble become antiparallel to each other, 
as is the case for nodal fermions \cite{rob}. 
Previous calculations show \cite{rob} that, as an interplay between the two effects,  the variation 
of $\omega_{sf}$ along the Fermi surface  near optimal doping is relatively modest, i.e, the 
crossover frequency for the normal state 
does not vary substantially along the Fermi surface.

\subsection{Superconducting state}

\subsubsection{polarization operator}
\label{pol_op}

We begin with the polarization operator, Eq.~(\ref{s2}).
Applying the spectral representation to (\ref{s2}), one can immediately see
 that in the superconducting state, $Im \Pi_{\bf q} (\Omega)$ remains 
linear in frequency only for $\Omega \gg \Delta$ (where 
$Im \Pi_{\bf q} (\Omega) = \gamma \Omega$). At smaller frequencies, 
$Im \Pi_{\bf q} (\Omega)$   vanishes below a threshold at 
$\Omega_{th} = |\Delta_{+}| + |\Delta_{-}|$, where, 
we remind, $\Delta_\pm = \Delta({\bf k}_m \pm {\bf q}/2)$, and
${\bf k}_m + {\bf q}/2$ and ${\bf k}_m - {\bf q}/2$ are discrete pairs of 
momenta (specified  by $m$), which are simultaneously on
 the Fermi surface. By 
 Kramers-Kronig relation, the vanishing of $Im \Pi_{\bf q} (\Omega)$ below the
 threshold generates a nonzero $Re \Pi_{\bf q} (\Omega)$, which comes from 
 frequencies of order $\Delta$ and is therefore part of the 
low-energy theory. This $Re \Pi_{\bf q} (\Omega)$  dominates the spin dynamics below the threshold. 

 The structure of $\Pi_{\bf q} (\Omega)$ in a $d-$wave superconductor 
 has been
 previously analyzed for ${\bf q} = {\bf q_0} = (\pi,\pi)$ \cite{res_peak_theory}. 
For ${\bf q} = {\bf q_0}$, different regions specified by $m$ are all equivalent, 
$\Delta ({\bf k}_F +{\bf q_0}) = 
-  \Delta ({\bf k}_F) = \Delta$, i.e., $\Omega_{th} = 2\Delta$. 
At the threshold frequency, 
 $Im \Pi_{{\bf q_0}} (\Omega)$ is discontinuous and 
jumps from zero to  $\pi \Delta \gamma$. 
By the KK relation, $Re \Pi_{{\bf q_0}} (\Omega)$
 then diverges  logarithmically at $\Omega_{th}$ \cite{ac}. 
This divergence 
 ensures that for arbitrary couping, 
 $\chi_s ({\bf q}, \Omega) = (\chi^{-1}_s ({\bf q}) - \Pi_{\bf q} (\Omega))^{-1}$ has an excitonic 
 pole at some $\omega_{res} < 2\Delta$, where 
$\Pi_{{\bf q_0}} (\omega_{res}) = Re \Pi_{{\bf q_0}} (\omega_{res}) = \chi^{-1}_s$.
At weak coupling, $\omega_{res}$ is exponentially 
 close to $2\Delta$, and the resonance is easily 
washed out by e.g.,  disorder. 
At strong coupling, however, the
 pole is located at  small frequencies and is weakly affected by disorder. 
Furthermore, expanding  
(\ref{s2}) in powers of $\Omega$, one can easily find 
 that at the lowest frequencies,
 $\Pi_{{\bf q_0}} (\Omega) \propto \Omega^2/\Delta$, i.e., at strong coupling, when 
$\omega_{res} \ll 2\Delta$, 
 the low-frequency spin susceptibility has a magnon-like form 
$\chi_s ({\bf q_0}, \Omega) \propto (\omega^2_{res} - \Omega^2)^{-1}$.   

The resonance behavior of $\Pi_{{\bf q_0}} (\Omega)$ sets the crossover 
 in the dispersion 
  of an antinodal fermion, for which the scattering by ${\bf q_0}$ 
is a low-energy process. For a nodal fermion, however, the 
scattering by ${\bf q_0}$ is ineffective, and one should analyze 
other ${\bf q}$ \cite{eschrig}. 
For a general ${\bf q} \neq {\bf q_0}$ connecting two Fermi surface points, 
 we find from Eq.~(\ref{s2}) 
that the magnitude of the jump in 
$Im \Pi_{\bf q} (\Omega)$ at the threshold frequency $\Omega_{th}
 = |\Delta ({\bf k}_F)|  + |\Delta ({\bf k}_F + {\bf q})|$ is
\begin{eqnarray}
&&\delta [Im \Pi_{\bf q} (\Omega_{th})] =
\frac{\pi \gamma}{2} \sqrt{|\Delta ({\bf k}_F)\Delta ({\bf k}_F + 
{\bf q})|} \nonumber \\
&&~\left(1- sgn (\Delta ({\bf k}_F)\Delta ({\bf k}_F + {\bf q}))\right)   
\label{s3}
\end{eqnarray}
It then follows that  for scattering from a
 nodal Fermi surface point ${\bf k}_{F,n}$ 
to some other point ${\bf k}_F = {\bf k}_{F,n} + {\bf q}$ 
along the Fermi surface, the
 jump in $Im \Pi_{\bf q} (\Omega_{th})$ disappears because 
$\Delta ( {\bf k}_{F,n}) =0$, 
even though $\Delta ({\bf k}_{F,n} + {\bf q}) \neq 0$. 
In the absence of a jump in $Im \Pi_{\bf q} (\Omega)$ at $\Omega_{th}$,
 $Re \Pi_{\bf q} (\Omega)$ does not diverge when
approaching $\Omega_{th}$ from below. Indeed, by the KK relation
\begin{equation}
Re \Pi_{\bf q} (\Omega) = \frac{2}{\pi}~\int_0^\infty 
\frac{dx Im \Pi_{\bf q} (x)}{x^2 - \Omega^2} 
\end{equation}
Near $\Omega = \Omega_{th}$, this reduces to
\begin{equation}
Re \Pi_{\bf q} (\Omega) \approx \frac{1}{\pi\Omega_{th}}
 \int_0^\infty \frac{dy Im \Pi_{\bf q} (y + \Omega_{th})}{y + 
(\Omega_{th} - \Omega)} 
\label{add1}
\end{equation}
When $Im \Pi_{\bf q} (\Omega_{th} + 0^+)$ has a non-zero value (which is the case when 
 $Im \Pi_{\bf q} (\Omega)$ jumps at the threshold), it can be pulled out from 
 the integral over $y$, and $Re \Pi_{\bf q} (\Omega_{th})$ diverges logarithmically.
Without the jump, $Im \Pi_{\bf q} (y + \Omega_{th})$ vanishes at $y=0^+$, and
 the integral over $y$ in (\ref{add1}) does not diverge. 
 We find from (\ref{s2})  that for scattering that involves a nodal fermion, 
 $Im \Pi_{\bf q} (y + \Omega_{th}) \propto y^{1/2}$, hence 
for $\Omega <\Omega_{th}$ we have from (\ref{add1}),  
\begin{equation}
Re \Pi_{\bf q} (\Omega) = - \gamma \sqrt{\Omega^2_{th} - \Omega^2}
\label{s33'}
\end{equation}
The minus sign in front of the square-root implies that $Re \Pi_{\bf q} (\Omega)$
 is negative, i.e., $\chi^{-1} ({\bf q}, \Omega) = \chi_s^{-1} ({\bf q}) - \Pi_{\bf q} (\Omega)$
does not change sign below $\Omega_{th}$, and the resonance mode does not 
emerge. We recall that a constant, $\Delta$ independent term, has been already 
pulled out from $Re \Pi_{\bf q} (\Omega)$, hence the negative value is with respect to 
the normal state (as $\chi^{-1}_s ({\bf q}) >0$ in a paramagnet, 
$\chi^{-1}_s ({\bf q}) -\Pi_{\bf q} (\Omega)$ is positive for all $\Omega < \Omega_{th}$).

Eq. (\ref{s33'}) can be easily extended to the full complex $ \Pi_{\bf q} (\Omega) $ which takes the form
\begin{equation}
\Pi_{\bf q} (\Omega) = i \gamma \sqrt{(\Omega + i \delta)^2
 - \Omega^2_{th}}
\label{s33}
\end{equation}
Note that  the 
square-root functional form of $\Pi_{\bf q} (\Omega)$ survives even when
the scattering is between a nodal fermion and a point not exactly on the Fermi surface.  
Indeed, as long as ${\bf q}$ is not directed 
along the zone diagonal,  one can set, without loss of generality,
 the velocity of a
 nodal fermion to be along the $y$ axis, i.e., $\epsilon_{\bf k}= y$, and 
 the velocity of a fermion at ${\bf k} +{\bf q}$ to be along $x$: 
$\epsilon_{{\bf k}+{\bf q}} = \epsilon_0 + x$ where $\epsilon_0 = \epsilon_{{\bf k}_F + {\bf  q}}$.
  Substituting this expansion into $\Pi_{\bf q} (\Omega) \propto \int 
G_{{\bf k}, \omega + \Omega} G_{{\bf k}+{\bf q}, \omega}$ we obtain
\begin{eqnarray}  
&&\Pi_{\bf q} (\Omega) \propto \int d \omega \frac{d x d y}{\omega + \Omega - y
 + i \delta_{\omega + \Omega}}\nonumber \\
&&~\frac{\epsilon_0 +x + \omega}{\omega^2 - (\epsilon_0 + x)^2 - \Delta^2 (x,y) + i \delta} 
\end{eqnarray}
Elementary analysis shows that the singular contribution to 
$\Pi_{\bf q} (\Omega)$ comes from $\omega = - \Omega$, $y =0$, and $x = - \epsilon_0$, 
i.e., from the internal momentum range when both fermions are back on the Fermi surface. 
Furthermore, as typical $y$ are infinitesimally small, one can neglect the $y$-dependence 
of the gap, in which case the momentum integral is factorized. Integrating over $y$, then over 
$\omega$, and finally over $x$, one immediately
 recovers (\ref{s33}) with $\Omega_{th} = \Delta (-\epsilon_0,0)$, that is the gap at the Fermi 
 surface point obtained by projecting 
${\bf k} + {\bf q}$ onto the Fermi surface along the
$x$ direction.

\begin{figure}[tbp]
\includegraphics[width=\columnwidth]{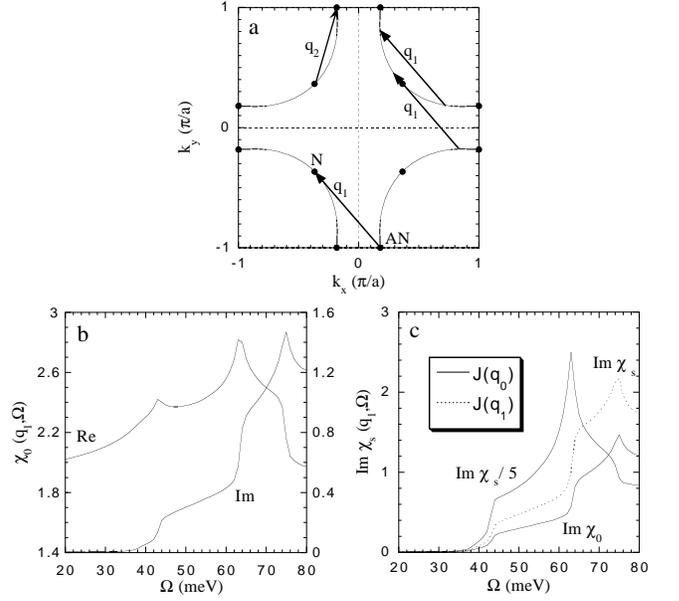}
\caption{a) Different pairs of Fermi surface points separated by 
 a momentum ${\bf q_1}$ equal to the node-antinode distance.  Solid points mark the node (N) and antinode (AN).
 In addition, the other node-antinode wavevector (${\bf q_2}$) is shown.  b)
 the imaginary and real parts of the particle-hole bubble $\chi_0({\bf q_1},\Omega)$.  The spin polarization
operator $\Pi({\bf q},\Omega)=\chi_0^{-1}({\bf q},0) - \chi_0^{-1}({\bf q},\Omega)$.
 Note near discontinuities in $Im\chi_0$ at $\Omega^\prime_{th} \approx 1.07 \Delta$ and $1.6 \Delta$.
 Here, $\Delta_{(\pi,0)}$=40 meV, $\delta$=0.5 meV, and $T$=0.5 meV.
 c) $Im \chi_s({\bf q_1}, \Omega)$ for two values of
 $J({\bf q})$ 
($\chi_s^{-1}({\bf q}) = \chi_0^{-1}({\bf q},0)-J({\bf q})$).  
Here, $J({\bf q_0})$ was chosen so as to yield
 a resonance at 40 meV for ${\bf q}={\bf q_0}$, and $J({\bf q_1})$ is $-J({\bf q_0}) (\cos(q_{1x}a)+\cos(q_{1y}a))/2$.}
\label{fig2}
\end{figure}

The square-root behavior of $\Pi_{\bf q} (\Omega)$ 
 is not the full story, however, as the same incommensurate ${\bf q}$
 which connects a nodal Fermi surface point with some other ${\bf k}_F$
may also connect other pairs of Fermi surface points 
for which the superconducting gap is non-zero for both points. If
 the signs of the two gaps are opposite, then, according to 
 (\ref{s3}),  $Im \Pi_{\bf q} (\Omega)$ still has discontinuities at 
the corresponding 
threshold frequencies $\Omega^\prime_{th}$, hence 
$Re \Pi_{\bf q} (\Omega^\prime_{th})$ diverges, and 
$\chi_s^{-1} ({\bf q}) -  Re \Pi_{\bf q} (\Omega)$ crosses zero at some frequency 
below $\Omega^\prime_{th}$. 
We checked this possibility for  the Fermi surface of
 optimally doped $Bi2212$ \cite{res_peak_theory}.
 For node-antinode scattering, there are two inequivalent ${\bf q}$ vectors.
The smaller of the two, which we label ${\bf q_2}$, has a dynamic
 response which is small relative to the larger of the two,
  which we label ${\bf q_1}$.  For ${\bf q_1}$, the node-antinode
 process at $\Omega_{th}=\Delta_{an}$ is the threshold ($an$ = antinode).  There are, though,
 two other inequivalent pairs of vectors for which 
  ${\bf k}$ and ${\bf k} + {\bf q_1}$ are both on the Fermi surface, 
 and $\Delta ({\bf k})$ and $\Delta ({\bf k} + {\bf q_1})$ 
have opposite signs. 
For these processes, $Im \Pi_{\bf q} (\Omega)$ 
is discontinuous at corresponding threshold frequencies $\Omega^\prime_{th} \approx 1.07 \Delta$ and $1.6 \Delta$.   
We illustrate this in Fig.~\ref{fig2}.
We found, however, that these extra processes do not give rise to 
a  resonance in the spin susceptibility for two reasons.
First, the dynamic response at ${\bf q_1}$ is weaker
than that at ${\bf q_0}$, and
the divergence in $Re \Pi_{\bf q_1}$ is further weakened at the
lower 
 energy $\Omega^\prime_{th}$ of the two since one of the two 
${\bf k}$ vectors is near the node in this case.  
Thus, the inclusion of any damping
(due to impurities or finite T) is enough to remove the divergence altogether.
Second, even in an idealized situation with zero fermionic damping, 
 the threshold frequencies $\Omega^\prime_{th}$ 
for both of these extra scattering  processes  exceed $\Omega_{th}$,
hence near $\Omega^\prime_{th}$, $Im \Pi_{\bf q_1} (\Omega)$ is already non-zero.
As a consequence, we find no resonance for $Im \chi_s({\bf q_1},\Omega)$
(though there can be a peak associated with the higher energy $\Omega^\prime_{th}$).

Summarizing, we argue that for ${\bf q_1}$ which connects nodal and antinodal Fermi surface
 points, there is no actual resonance in the spin susceptibility.
The imaginary part of $\chi_s ({\bf q},\Omega)$ emerges at $\Omega_{th} = \Delta_{an}$ as 
$(\Omega - \Delta_{an})^{1/2}$ and has extra bumps at higher energies near threshold frequencies for
 additional scattering processes  with the same ${\bf q}$.
Alternatively speaking, the excitonic resonance in the dynamical spin susceptibility exists for ${\bf q}$ 
near ${\bf q_0}$ but gradually vanishes as ${\bf q}$ approaches ${\bf q_1}$.  The boundary between these
two regions is roughly set by the diagonal node-node scattering vector, which has a length intermediate
between ${\bf q_0}$ and ${\bf q_1}$.

\subsubsection{fermionic self-energy}

We next compute the fermionic  self-energy, Eq.~(\ref{s1}).  
For an antinodal fermion, both ${\bf k}$ and ${\bf k+q_0}$ are near the Fermi surface,
 and the resonance mode has a strong effect on the self-energy. 
Assuming that $\chi_s ({\bf q}, \Omega)$ has a magnon-like form 
 $\chi_s ({\bf q}, \Omega) = \chi_0/ (\omega^2_{res}({\bf q}) - \Omega^2)$, adding a small 
 damping term $i\delta$  to $\Omega$ for continuity, and 
neglecting  the momentum dependence of $\omega_{res}$ both for simplicity and 
 because  experimentally, the resonance dispersion is rather flat \cite{bourges},
 we obtain from (\ref{s1}) in the superconducting state
\begin{eqnarray}
\Sigma_{SC,an} (\omega) &=& \frac{3 i g^2_s \chi_0}{4 \pi^2 v_F}~\int 
\frac{dx d \Omega (\omega + \Omega +x)}{x^2 + \Delta^2 - (\omega + \Omega)^2 - i \delta}~\nonumber \\
&&\frac{1}{\omega^2_{res} - \Omega^2 - i \delta}
\label{z1}
\end{eqnarray}
The subscript for the self-energy implies 
 SC = superconducting state, an = antinode.
 We also defined $x = v_F q_{\perp}$ and used the superconducting Green's function for free fermions 
\begin{equation}
G^0_{SC} ({\bf k}, \omega) = \frac{\omega + \epsilon_{\bf k}}{\omega^2 + i \delta - \Delta^2 - \epsilon^2_{\bf k}}
\label{z2}
\end{equation}
The integration over $x$ is straightforward, and performing it we obtain
\begin{eqnarray}
\Sigma_{SC,an} (\omega) &=& -\frac{3  g^2_s \chi_0}{4 \pi v_F}~
\int \frac{d \Omega}{\omega^2_{res} - \Omega^2 - i \delta}~\nonumber 
\\
&&\frac{\omega + \Omega}{\sqrt{ (\omega + \Omega)^2 - \Delta^2 + i \delta}}
\label{z3}
\end{eqnarray}
This integral is singular near 
 $\omega = - \omega_0 = -( \Delta + \omega_{res})$. To see this, 
 introduce $\omega = - \omega_0 + \epsilon$, and $\Omega = \omega_{res} + y$.
Substituting these expansions into (\ref{z3}) and restricting to only linear terms in $y$ and $\epsilon$, 
we obtain after simple algebra
\begin{equation}
 \Sigma_{SC,an} (\epsilon) = -\frac{3  g^2_s \chi_0 \sqrt{\Delta}}{8 \pi \sqrt{2}
\omega_{res} v_F}~\int \frac{dy}{y + i \delta} \frac{1}{\sqrt{- \epsilon - y + i \delta}}
\label{z4}
\end{equation}
Splitting the integration over $y$ into integrals over 
positive and negative $y$ and evaluating them separately, we obtain 
\begin{equation}
\int_{-\infty}^\infty \frac{dy}{y + i\delta}~\frac{1}{\sqrt{-\epsilon-y + i \delta}} =
 - \frac{2\pi}{\sqrt{\epsilon-i\delta}}
\label{ops}
\end{equation}
Substituting this into (\ref{z4}) we obtain, 
\begin{equation}
 \Sigma_{SC,an} (\epsilon) = \frac{3  g^2_s \chi_0 \sqrt{\Delta}}{8\omega_{res} v_F} \frac{\sqrt{2}}{\sqrt{\epsilon - i \delta}}
\label{z5}
\end{equation}
Separating real and imaginary parts of $1/\sqrt{\epsilon - i \delta}$
 and replacing $\epsilon$ back by $\omega + \omega_0$ we finally obtain 
\begin{equation}
\Sigma^\prime_{SC,an} (\omega) = \lambda_{eff} ~\Delta^{3/2}
 \frac{(\omega +\omega_0 + 
\sqrt{(\omega_0 + \omega)^2 + \delta^2})^{1/2}}{\sqrt{(\omega_0 + \omega)^2 + 
\delta^2}}
\label{s4}
\end{equation}
where 
\begin{equation}
\lambda_{eff} = \frac{3 g_s^2 \chi_0}{8 v_F \Delta^2}~\frac{\Delta}{\omega_{res}}
\label{z6}
\end{equation}
Note that $\lambda_{eff}$ is dimensionless 
(with the above definition of  $\chi_0$).
  
\begin{figure}[tbp]
\includegraphics[width=\columnwidth]{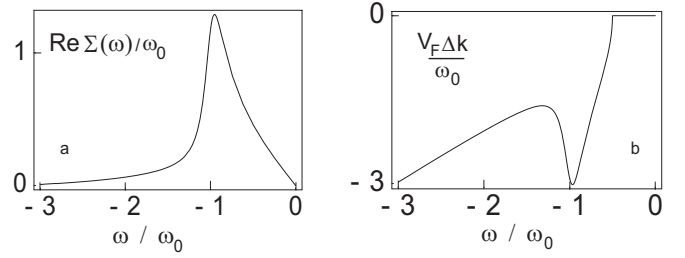}
\caption{The self-energy and the dispersion 
 near an antinodal point. The coupling is 
$\lambda ({\bf q_1}) =2$. We also set 
 $\Delta = \omega_{res}$, i.e., $\omega_0 = 2 \Delta$, and 
 use a broadening $\delta = 0.3 \Delta$.
Note the $S-$shape dispersion near $\omega_0$. }
\label{fig3a}
\end{figure}

 We see from (\ref{s4})
 that the real part of the self-energy has a near one-sided singularity. It 
almost diverges as $1/\sqrt{\omega_0 + \omega}$ as $\omega$  approaches 
$-\omega_0$ from above ($|\omega| < \omega_0$), and then rapidly drops 
beyond $-\omega_0$, reducing to $O(\delta^2/(\omega + \omega_0))$ 
 when $|\omega| > \omega_0$. 
The relation between this self-energy and the dispersion is somewhat 
complicated in a superconductor, as the MDC lineshape does not 
have a simple lorentzian form because the Green's function has 
$\epsilon_{\bf k} = v_F (k_\perp -k_F)$ both in the numerator and the denominator:
\begin{equation}
G({\bf k}, \omega) = \frac{\omega - \Sigma (\omega) + \epsilon_{\bf k}}{(\omega - \Sigma (\omega))^2 - 
\Phi^2 (k_{\parallel}, \omega) - \epsilon^2_{\bf k}}
\label{feb3_1}
\end{equation}
Here  $\Phi (k_{\parallel}, \omega)$ is the pairing  
vertex. It is related to the pairing gap 
 $\Delta (k_{\parallel}, \omega)$ by $\Phi (k_{\parallel}, 
\omega) = \Delta (k_{\parallel}, \omega) Z (\omega)$, where
$Z(\omega) =  1 - \Sigma (\omega)/ \omega$~~\cite{eschrig,acs2,carbotte}.
For simplicity, we neglect the frequency dependence 
 of $\Delta (k_{\parallel}, \omega)$, i.e., approximate  
$\Delta (k_{\parallel}, \omega)$ by a frequency independent gap $\Delta (k_{\parallel})$. 
Near the antinodal points, the gap is near its maximum, i.e., is rather flat as a function of $k_{\parallel}$, and can be
 approximated by a constant $\Delta$. 
Still, the presence of $\epsilon_{\bf k}$ in the numerator 
 of (\ref{feb3_1}) 
implies that the maximum of 
$Im G({\bf k}, \omega)$ is
 shifted somewhat in $\omega$
from where the real part of the the denominator in (\ref{feb3_1}) vanishes 
(in the BCS limit, this effect can be attributed to the $k-$dependence of the coherence factors).
This complication, however, affects the form of the dispersion mainly for $|\omega| <  \Delta$, and is less 
relevant near $\omega = - \omega_0$ where the self-energy is nearly singular.
To avoid this complication, we neglect the $k-$dependence of the numerator of the Green's function, 
and extract the dispersion from
\begin{equation}
Re \left[\omega - \Sigma (\omega)  + \sqrt{\Phi^2 (\omega) + v^2_F (k_\perp -k_F)^2}\right] =0
\label{feb2_1}
\end{equation}
Substituting $\Phi (\omega)$ in terms of $\Delta$, and neglecting 
 $Im Z(\omega)$ (which vanishes for $|\omega| < \omega_0$ anyway), we 
 obtain from (\ref{feb2_1})
\begin{equation}
(\omega - \Sigma^\prime_{SC,an} (\omega))~Re 
\sqrt{\frac{\omega^2 - \Delta^2}{\omega^2}} = v_F (k_\perp -k_F)
\label{feb2_2}
\end{equation}     
Substituting $\Sigma^\prime_{SC,an}$ from (\ref{s4}), we find that 
 the dispersion 
develops an $S$-shape 
for $|\omega| < \omega_0$, precisely as seen in the experiments. 
We illustrate this in Fig.~\ref{fig3a}.
We also recall that 
 a near divergence of $Re \Sigma_{an} (\omega_0)$ implies, by KK transform, 
 a near discontinuity in   $Im \Sigma_{an}
 (\omega_0)$, both of which  
give rise to the experimentally observed
peak/dip/hump behavior of the ARPES intensity  \cite{pdh,ding}.

For a nodal fermion, the situation is different. A shift by 
${\bf q_0}$ moves a nodal Fermi surface point to a point where the
 energy is very large, $\sim 0.7 eV$ for optimally doped $Bi2212$. 
The $(\pi,\pi)$ scattering is then ineffective. 
However, for ${\bf q}$ which we analyzed in the previous subsection, 
 a nodal fermion can still scatter along the 
 Fermi surface, which gives rise to a much larger self-energy.
  
The computation of the self-energy $\Sigma_{SC,n} (\omega)$ requires
 extra care, as we will have to average over all ${\bf q}$ which connect a nodal point 
 with other points on the Fermi surface. Besides, even for the integration near a particular ${\bf q}$,  
the dispersion of the $d-$wave gap is essential
 as it affects the functional form of 
 $Im \Sigma_{SC,n} (\omega)$ via the softening of the singularity in the 
 fermionic  density of states at fermionic frequencies near the gap
 at ${\bf k}_F + {\bf q}$. 
As our goal is to 
demonstrate that  the self-energy at the nodal point does not have the sharp 
features of the antinodal self-energy, we assume for simplicity that (i) the
 dominant contribution to $\Sigma_{SC,n} (\omega)$ comes from node-antinode scattering,
 because of the presence of the density of states singularity associated with the antinode,
and (ii) that the superconducting gap has a flat dispersion in the antinode region. 

\begin{figure}[tbp]
\includegraphics[width=\columnwidth]{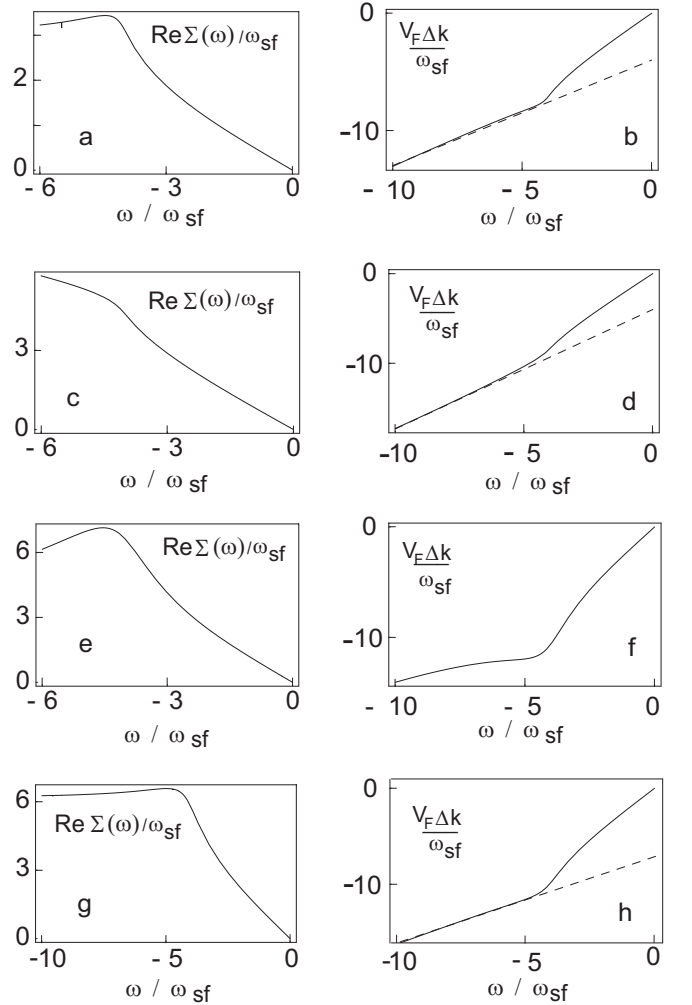}
\caption{$Re\Sigma$ and the dispersion 
 along the nodal direction. a)-f) are obtained with $\Pi_{\bf q} (\omega)$ 
given by (\protect\ref{s33'}). 
a)-b)  are for a flat static spin susceptibility  $\chi_s ({\bf q})$ near ${\bf q_1}$, c)-h) for 
the Ornstein-Zernike form. c)-d) are obtained within Eliashberg theory,
e)-f) assuming a
vanishing Fermi velocity at ${\bf q}_1$ (anti-Eliashberg approximation). 
g) and h)
 are obtained using the spin polarization operator, Eq.~(\protect\ref{dec_4}),
 with two extra thresholds at $\Omega^\prime_{th}$.
In all cases, the dispersion 
shows a kink around $2\Delta$, and does not have the $S-$shape form typical for the 
antinodal dispersion (see Fig.~\protect\ref{fig3a}). 
Dashed lines are extrapolations from high frequencies; the extrapolated dispersion 
at $\omega=0$ has a negative $\Delta k$, i.e., $k$ inside the Fermi surface. 
The coupling $\lambda ({\bf q_1}) =2$ and $\Delta/\omega_{sf} ({\bf q_1}) =2$.}
\label{fig4}
\end{figure}

 Substituting     
the spin susceptibility with  $\Pi$ given by (\ref{s33}) 
into Eq.~(\ref{s1}) and  neglecting 
 momentarily the  dispersion in $\chi_s ({\bf q})$ around ${\bf q_1}$, 
we obtain after 
integrating over momentum near the Fermi surface ($\Omega_{th}=\Delta$)  
\begin{eqnarray} 
\Sigma_{SC,n} (\omega) &=& \frac{\lambda ({\bf q_1})}{2}~ 
 \int_{-\infty}^{\infty} d \omega^\prime 
\frac{\omega^\prime}{\sqrt{(\omega^\prime)^2 - \Delta^2}}~\times \nonumber \\
&&\frac{1}{1 - 
i \sqrt{(\omega + \omega^\prime)^2 -\Delta^2}/\omega_{sf} ({\bf q_1})}
\label{s5}
\end{eqnarray}
where $\omega_{sf} ({\bf q_1}) = (\gamma_{{\bf q_1}} \chi_s ({\bf q_1}))^{-1}$.
At the lowest frequencies,
expanding to linear order in $\omega$ and collecting the prefactor, 
 we obtain 
\begin{equation}
Re \Sigma_{SC,n} (\omega) = -\lambda_{sc} ({\bf q_1}) \omega,
\end{equation}
 where
$\lambda_{sc} ({\bf q_1}) = \lambda ({\bf q_1})~ \psi (\Delta/\omega_{sf} ({\bf q_1}))$,
 and 
\begin{equation}
\psi (x) = \int_0^\infty \frac{dz}{(z^2 +1)^{3/2}}~\frac{1}{1 + x \sqrt{z^2 +1}}
\end{equation}
such that $\psi (x) \leq  \psi (0) =1$. This implies 
that the coupling constant in the superconducting state is 
 somewhat smaller than in the normal state. 
 This is in agreement with earlier work \cite{acs}. 
At larger frequencies, 
$Re \Sigma_{SC,n} (\omega)$ is continuous and  reduces to its 
nearly flat  normal state form, Eq.~(\ref{need1}) at $|\omega| >> \Delta, 
\omega_0$. The limiting behavior  resembles that in 
 the  normal state, however  
the crossover in (\ref{s5}) 
is not analytic, and the self-energy develops a kink at 
$\omega = -2\Delta$. This can be most easily seen by evaluating the derivative
 of the self-energy. Indeed, differentiating with respect to $\omega$ in (\ref{s5}), we obtain 
\begin{eqnarray}
\frac{d \Sigma_{SC,n} (\omega)}{d\omega} &=&
 \frac{i \lambda ({\bf q_1})}{2 \omega_{sf} ({\bf q_1})} \int 
\frac{d\omega^\prime}{\sqrt{(\omega^\prime)^2-\Delta^2} 
\sqrt{(\omega + \omega^\prime)^2 -\Delta^2}}\nonumber \\
&&~\frac{\omega^\prime (\omega + \omega^\prime)}{(1 - 
i \sqrt{(\omega + \omega^\prime)^2 -\Delta^2}/\omega_{sf} ({\bf q_1}))^2}
\label{dec1_1}
\end{eqnarray}
Near $\omega = \pm 2\Delta$, the dominant contribution to the integral comes from 
$\omega^\prime \approx -\Delta sgn (\omega)$, when the two branch-cut singularities 
merge. In this region,
 $\omega^\prime (\omega + \omega^\prime) \approx -\Delta^2$, 
 $(\omega + \omega^\prime)^2 \approx \Delta^2$, and 
\begin{eqnarray}
&&\frac{d \Sigma_{SC,n} (\omega)}{d\omega} \approx \nonumber \\
&& \frac{-i \lambda ({\bf q}_1) \Delta^2}{2 \omega_{sf} ({\bf q}_1)} \int 
\frac{d\omega^\prime}{\sqrt{(\omega^\prime)^2-\Delta^2} 
\sqrt{(\omega + \omega^\prime)^2 -\Delta^2}}
\label{dec1_2}
\end{eqnarray}   
Evaluating the integral, we find that at, e.g. $\omega = -2 \Delta$, 
$d Im \Sigma (\omega)/d \omega$ undergoes a finite jump 
\begin{eqnarray}
&&\frac{d}{d \omega}  \left[
Im \Sigma_{SC,n} (-2 \Delta - \epsilon) - Im \Sigma_{SC,n}
(-2\Delta +\epsilon)\right] \propto \nonumber \\
&& \int_0^{\epsilon} 
\frac{dx}{\sqrt{x} \sqrt{\epsilon -x}} = 2 \int_0^1 \frac{dz}{\sqrt{1-z^2}} =\pi \label{dec1}
\end{eqnarray}
By the KK relation, $d Re \Sigma (\omega)/d \omega$  diverges logarithmically 
 at $\omega = - 2\Delta$.  
This behavior is analogous to the behavior of the polarization operator
 with ${\bf q} = (\pi,\pi)$ near the threshold frequency $\Omega_{th} = 2\Delta$. 
Evaluating $d\Sigma^\prime_{SC,n} (\omega)/d \omega$  explicitly,
introducing $\omega = - 2\Delta + 
\epsilon$ and $\omega^\prime = \Delta + x$, and expanding to first order in $x$ and in 
$\epsilon$ in the two terms in the denominator, we obtain near $\omega = -2\Delta$
\begin{equation}
\frac{d \Sigma^{\prime}_{SC,n} (\omega)}{d\omega} \approx \frac{\lambda ({\bf q}_1) \Delta}
{4 \omega_{sf}  ({\bf q}_1)}~Im \left[\int_{-A}^{A} \frac{dx}{\sqrt{x+ i \delta}\sqrt{- \epsilon - x + i \delta}}\right]
\label{z7}
\end{equation}
Here $A \sim \Delta$ is the upper cutoff for the linear expansion.
Splitting the integral into the integrals over positive $x$ and over negative $x$, and evaluating 
them separately, we obtain, to logarithmical accuracy
\begin{equation}   
\frac{d \Sigma^{\prime}_{SC,n} (\omega)}{d\omega} 
\approx -K~\log{\frac{\Delta}{|\omega + 2 \Delta|}}
\label{z8}
\end{equation}
where $K = \lambda ({\bf q_1}) \Delta/(2\omega_{sf} ({\bf q_1}))$.
Integrating this formula, we obtain
\begin{equation}
\Sigma^\prime_{SC,n} (\omega) = \Sigma^\prime_{SC,n} (-2\Delta) 
- K~ (\omega +2\Delta)
 \log{\frac{\Delta}{|\omega +2\Delta|}}
\label{s7}
\end{equation} 
Note that, contrary to (\ref{s4}), the singularity in 
$\Sigma^\prime_{SC,n} (\omega)$ 
is two-sided, i.e., is symmetric with respect to $\omega + 2\Delta$. 
 Substituting this self-energy into the dispersion of a nodal fermion 
 $\omega - \Sigma^{\prime} (\omega) = v_F (k_{\perp} - k_F)$, 
 we find that the kink in $\Sigma^\prime$ 
gives rise to a kink in the dispersion at $\omega = - 2 \Delta$, but 
 the $S$-shape form {\it does not emerge}. We plot $Re \Sigma (\omega)$ and 
 the dispersion in  Figs.~\ref{fig4}a and \ref{fig4}b.

The inclusion of the momentum dispersion of $\chi_s ({\bf q})$ around ${\bf q_1}$ 
 affects the prefactor, but the logarithmical singularity near $-2\Delta$ survives.  
 Indeed,  this singularity is just  the consequence of the 
 square-root non-analyticity in the spin polarization operator, which is confined to  
 bosonic frequencies near $\Delta$. At these frequencies, 
 $\Pi_{{\bf q_1}} (\omega)$ is small, and the
 dynamical spin susceptibility can be  expanded in powers of $\Pi_{{\bf q_1}} (\omega)$. The momentum
 dependence of $\chi_s ({\bf q})$  then just affects the prefactor of the 
 $(\omega + 2\Delta) \log (\omega +2 \Delta)$ term in (\ref{s7}). 
 To verify  this, we  computed the self-energy and the dispersion for the
 Ornstein-Zernike form of $\chi_s ({\bf q})$. In this situation, 
\begin{eqnarray} 
&&\Sigma_{SC,n} (\omega) = \frac{\lambda ({\bf q_1})}{2}~ 
 \int_{-\infty}^{\infty} d \omega^\prime 
\frac{\omega^\prime}{\sqrt{(\omega^\prime)^2 - \Delta^2}}~\times \nonumber \\
&&\frac{1}{(1 - 
i \sqrt{(\omega + \omega^\prime)^2 -\Delta^2}/\omega_{sf} ({\bf q_1}))^{1/2}}
\label{s5_1}
\end{eqnarray}
Again, at low frequencies
\begin{equation}
Re \Sigma_{SC,n} (\omega) = -\lambda_{sc} ({\bf q_1}) \omega,
\end{equation}
 where now 
$\lambda_{sc} ({\bf q_1}) = \lambda ({\bf q_1})~ {\tilde \psi} 
(\Delta/\omega_{sf} ({\bf q_1}))$,
 and 
\begin{equation}
{\tilde \psi} (x) = \int_0^\infty \frac{dz}{(z^2 +1)^{3/2}}~
\frac{1}{(1 + x \sqrt{z^2 +1})^{1/2}}
\end{equation}
Near $\omega = -2\Delta$, we obtain from (\ref{s5_1}) the same functional form as in (\ref{s7})
\begin{equation}
\Sigma^\prime_{SC,n} (\omega) = \Sigma^\prime_{SC,n} (-2\Delta) 
- {\tilde K}~ (\omega +2\Delta)
 \log{\frac{\Delta}{|\omega +2\Delta|}}
\label{s7_1}
\end{equation} 
where now  ${\tilde K} = \lambda ({\bf q_1}) \Delta/(4\omega_{sf} ({\bf q_1}))$.
We plot the self-energy and the dispersion for the Ornstein-Zernike static susceptibility 
in Figs.~\ref{fig4}c and \ref{fig4}d. We see that the dispersion does not change much 
 from that for a flat $\chi_s ({\bf q})$, namely there is a kink near $2\Delta$, but 
 there is no $S-$shape dispersion as occurs for an antinodal fermion.  

Finally, to verify that the nodal dispersion is not an artifact of our computational procedure, 
we also  computed $\Sigma^\prime_{SC,n} (\omega)$ with the
Ornstein-Zernike $\chi_s ({\bf q})$ in a different (anti-Eliashberg)
  computational scheme:
 we assumed  that the Fermi velocity of an antinodal fermion 
nearly vanishes, and integrated over momenta normal to the Fermi surface in the bosonic 
rather than the fermionic propagator.
In this situation, 
\begin{eqnarray} 
\Sigma_{SC,n} (\omega) &=& \frac{-i\Delta\lambda ({\bf q_1})}{2}~ 
 \int_{-\infty}^{\infty} d \omega^\prime 
\frac{\omega^\prime}{(\omega^\prime)^2 - \Delta^2}~\nonumber \\
&&\log{\frac{
\omega_{sf}({\bf q_1}) - i \sqrt{(\omega + \omega^\prime)^2 -\Delta^2}}
{\omega_{sf}({\bf q_1}) - i \sqrt{(\omega - \omega^\prime)^2 -\Delta^2}}}
\label{s5_2}
\end{eqnarray}
Again, at low frequencies, the self-energy is linear,
\begin{equation}
Re \Sigma_{SC,n} (\omega) = -\lambda_{sc} ({\bf q_1}) \omega,
\end{equation}
 where now 
$\lambda_{sc} = \lambda ({\bf q_1}) \psi^* (\Delta/\omega_{sf} ({\bf q_1}))$, and 
$\psi^* (x << 1) = x |\log x|$, and $\psi^* (x >>1) = \pi/4$.
Near $\omega = \pm 2 \Delta$,
 the derivative $d \Sigma_{SC,n} (\omega)/d\omega$ is again singular, but now
\begin{eqnarray}
&&\frac{d \Sigma_{SC,n} (\omega)}{d\omega}  = - \Delta \lambda ({\bf q}_1) 
\int_{-\infty}^\infty d \omega^\prime 
\frac{\omega^\prime (\omega + \omega^\prime)}{((\omega^\prime)^2 -
 \Delta^2 + i \delta)}\times \nonumber \\
&& \frac{1}{ \sqrt{(\omega + \omega^\prime)^2 -\Delta^2 + i \delta}}
~\frac{1}{\omega_{sf} ({\bf q}_1) - i \sqrt{(\omega + \omega^\prime)^2 - \Delta^2}} \nonumber \\
\label{z9}
\end{eqnarray}
Expanding, as before near $\omega =-2\Delta$ and introducing 
$\omega = -2\Delta + \epsilon$ and $\omega^\prime = \Delta +x$, we obtain from (\ref{z9})
\begin{equation}
 \frac{d \Sigma_{SC,n} (\omega)}{d\omega} = \frac{\Delta^2 \lambda ({\bf q}_1)}
 {2 \sqrt{2\Delta} \omega_{sf} ({\bf q}_1)} \int_{-\infty}^\infty \frac{dx}{x + i \delta}~\frac{1}{\sqrt{-\epsilon -x + i\delta}}
\label{z10}
\end{equation}
Evaluating the integral using  (\ref{ops}) we obtain 
\begin{equation}
\frac{d \Sigma_{SC,n} (\omega)}{d\omega} = -\frac{\pi \Delta^2 
\lambda ({\bf q}_1)}{\sqrt{2\Delta} \omega_{sf} ({\bf q}_1)}~\frac{1}{\sqrt{\omega + 2 \Delta}}
\end{equation}
i.e, near $\omega = -2\Delta$,
\begin{equation}
\Sigma^\prime_{SC,n} (\omega) = \Sigma^\prime_{SC,n} (-2\Delta) 
- K^* Re \sqrt{2\Delta (\omega +2\Delta)}
\end{equation}
where $K^* = \pi \lambda ({\bf q}_1) \Delta/(\omega_{sf} ({\bf q}_1))$. 
As a result, the kink at $-2\Delta$ survives, and  
 $\Sigma_{SC,n} (-2\Delta)$ still does not diverge,
 i.e., there is no $S-$shape, antinodal-type dispersion.
  At the same time, 
  the functional form of the non-analytic piece changes --
 the $x \log |x|$ singularity in the real part of the 
self-energy in (\ref{s7}) and (\ref{s7_1}),
where $x = \omega + 2\Delta$, gets replaced by a 
one-sided $\sqrt{x}$ singularity.
We plot the self-energy, Eq.~(\ref{s5_2}), and the resulting dispersion
 in Figs.~\ref{fig4}e and \ref{fig4}f.  As the two 
expressions for the self-energy for the
Ornstein-Zernike $\chi_s ({\bf q})$, Eqs.~(\ref{s5_1}) and (\ref{s5_2}), 
 represent two extremes of large and small fermionic dispersion compared to 
the bosonic dispersion,
 the actual self-energy should be somewhere in between, i.e., it is 
 stronger than in (\ref{s7}) for $|\omega| <2\Delta$, and 
 weaker for $|\omega|>2\Delta$. Still, we emphasize that both computational schemes
 yield a kink in the dispersion, but no divergence of $Re \Sigma (-2\Delta)$ and no $S-$shape  dispersion. 
  
The actual behavior of the dispersion is more involved, as
 one has to average over all ${\bf q}$ for scattering from a node to some other 
Fermi surface point. This obviously weakens the 
$2\Delta$ singularity 
 roughly in the same way as the singularity in the density of states at $\Delta$ is weakened
 by the momentum dependence of the gap in a $d-$wave superconductor. 
Furthermore, even if by geometrical reasons, 
 node-antinode scattering at ${\bf q_1}$ dominates the nodal 
self-energy, the actual form of $\Sigma_{SC,n} (\omega)$ is more complex than in 
(\ref{s5}), (\ref{s5_1}) or (\ref{s5_2}).  As we already discussed 
in Sec.~\ref{pol_op},
  the square-root behavior of $\Pi_{{\bf q}} (\Omega)$ persists only in a small region near the threshold, while at 
 larger $\Omega$, additional features in $\Pi_{{\bf q_1}} (\Omega)$,
 associated with the existence of two extra 
pairs of Fermi surface points separated by ${\bf q_1}$ become relevant 
(see  Fig.~\ref{fig2}). 

To qualitatively estimate the relevance of this effect, we 
model 
$\Pi_{{\bf q_1}} (\Omega)$ in Fig.~\ref{fig2} by a combination of a square-root behavior above 
$\Omega =\Delta$ and a near discontinuity at $\Omega = 1.6 \Delta$: 
\begin{equation}
\Pi^{\prime \prime}_{{\bf q_1}} (\Omega) \approx 
 \gamma \left(\sqrt{\Omega^2 - \Delta^2} + 
\frac{a}{2} (1 + \tanh{\frac{\Omega^2 - (1.6\Delta)^2}{b}})\right)
\label{dec_4}
\end{equation}
For consistency with Fig.~\ref{fig2}, $a \sim 1.2\Delta$, and we set $b \sim 0.1 \Delta$ 
($b=0$ would correspond to a true discontinuity). 
In Figs.~\ref{fig4}g and \ref{fig4}h, we plot the 
self-energy and the dispersion for 
this form of $\Pi_{\bf q} (\Omega)$ and  
flat $\chi_s ({\bf q})$. We see that the dispersion does not 
change much compared to  that with just a square-root form of 
$\Pi_{{\bf q_1}} (\Omega)$ (see Fig.~\ref{fig4}b). There is still a cusp 
near $2\Delta$ and a non $S-$shape dispersion. The only new effect is 
the extension of the crossover  region above the kink. This is indeed
 expected as the new term in (\ref{dec_4}) 
affects the polarization operator at higher frequencies. 

To summarize this subsection, we  see that the result of various computational procedures 
is virtually the same:
 the self-energy of a nodal fermion is much less affected by 
superconductivity than the self-energy of an antinodal fermion. 
The nodal self-energy roughly displays the same crossover as in the normal state, 
from a linear in frequency behavior at small frequencies, to a more 
flat behavior at higher frequencies. Superconductivity only sharpens the crossover near 
$2\Delta$, but does not give rise to any $S-$shape features in the dispersion.
       
\begin{figure}[tbp]
\includegraphics[width=\columnwidth]{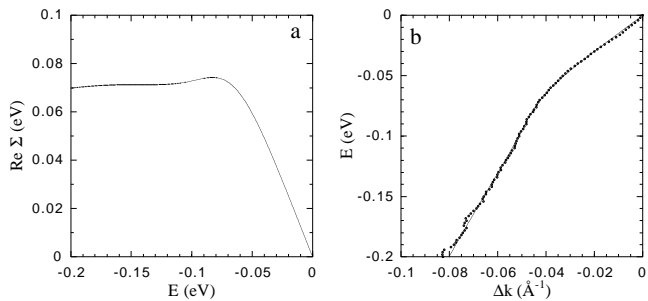}
\caption{a) Re $\Sigma$ and b) nodal dispersion generated from the model of Ref.~\onlinecite{eschrig}.
Circles are the experimental data of Ref.~\onlinecite{kaminski} for optimal doped Bi2212 at T=40K.
The S-shape is replaced by a kink due to (1) dominance of the continuum contribution
to the self-energy relative to that of the  resonance and (2) a strong
energy broadening of 40 meV which was assumed when deriving the self-energy.
For these calculations, $\omega_{res}$=40 meV, $\Delta_{(\pi,0)}$=40 meV, and
the continuum has a gap of 65 meV and a cut-off of 500 meV.  
For this direction, the bare Fermi
velocity is 3.37 eV$\AA$.  }
\label{fig5}
\end{figure}

\subsection{a comparison with Ref.~\onlinecite{eschrig}}
 
It is instructive to compare our results for nodal fermions 
with the earlier study by Eschrig and Norman \cite{eschrig}. These 
 authors also argued that the dominant scattering process for a nodal 
fermion is  node-antinode scattering by ${\bf q_1}$ (though they included
a second process due to node-$(\pi,0)$ scattering). They used a 
phenomenological form of the spin susceptibility at ${\bf q_1}$ with
the resonance piece taken as a Lorenztian of width 2$a$ about
${\bf q_0}$.  The resulting value at ${\bf q_1}$ is about 8\% of that
at ${\bf q_0}$.
Added to this is a gapped continuum (with a gap equal to the threshold value of
$Im \Pi_{{\bf q_0}}$) modeled as a step function with an ultraviolet
cut-off.
This form of $\chi_s ({\bf q_1}, \Omega)$ is not exactly the one we used above
 (in our analysis, the resonance piece is completely absent at ${\bf q_1}$), but 
 nevertheless is rather similar in the sense that a large part of the
 magnetic excitation spectrum is incoherent. 
 Not surprisingly, the two forms for $\chi_s ({\bf q_1}, \omega)$ yield 
similar results for the dispersion of a nodal electron.
In Fig.~\ref{fig5}, we plot the self-energy and nodal dispersion obtained with their
susceptibility. We see that this dispersion has a kink, 
but no $S-$shape. This was achieved by putting in a significantly large amount of
damping which acts to smear out the delta function singularity in energy of the
resonance.  An $S$-shape would still be present, though, if the continuum piece
is ignored.  This result shows that even when the resonance is 
present, the $S-$shape of the dispersion emerges only 
 when the resonance contribution to $\Sigma^\prime (\omega)$ overshadows the one from the  
 incoherent piece in $\chi_s ({\bf q}, \Omega)$.
This implies, in particular, that the $S-$shape dispersion does not emerge immediately away 
from the nodal direction, i.e., it appears somewhere between the nodal and antinodal points.

\begin{figure}[tbp]
\includegraphics[width=\columnwidth]{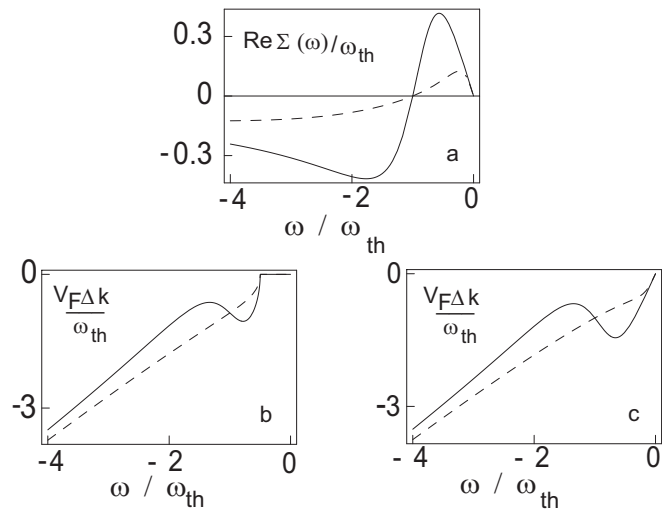}
\caption{a) $Re\Sigma$ and b) antinodal and c) nodal
dispersion for the electron-phonon interaction. We used Eq.~(\protect\ref{4})
 for the self-energy with
 $\omega_{th} = 2 \Delta$ and coupling 
 constant $\lambda^{ep}_{sc} =2$. Solid and dashed lines are for 
 $\gamma/\omega_{th} = 0.6$ and $2$, respectively. 
For moderate damping, the dispersion along both directions  has an 
$S-$shape form. 
Note that the extrapolated dispersion 
crosses the vertical axis at a positive 
$\Delta k$, i.e., for $k$ outside the Fermi surface.
 For the spin-fermion interaction, this crossing always occurs
 at negative $\Delta k$ for the nodal direction.}
\label{fig6}
\end{figure}

\section{ Phonon scattering}

We now perform the same analysis for phonon scattering. 
Consider a system of electrons interacting with an optical phonon with a frequency 
$\omega_{ph}$ and momentum ${\bf q}={\bf q}_{ph}$. The optical phonon propagator  
 can be reasonably approximated by
\begin{equation}
\chi_{ph} ({\bf q}, \omega) = \frac{g({\bf q})}{(\omega + i \gamma)^2 - \omega^2_{ph}}
\label{p3}
\end{equation}
where $g({\bf q})$ rapidly decays at deviations from ${\bf q}_{ph}$, and  $\gamma$ 
is the phonon damping rate.
The fermionic self-energy  due to electron-phonon scattering is given by (\ref{s1}), 
only $3 g_s^2$ is replaced by $g^2_{ep}$.
To simplify the discussion, we consider a nodal fermion and assume 
that the Fermi velocity at ${\bf k}_F + {\bf q}_{ph}$ can be neglected.
Substituting (\ref{p3}) into the self-energy, we obtain
\begin{equation}
Re \left[\Sigma^{ph}_{nodal} (\omega) \right]= - \frac{\pi \beta \omega}{\omega_{ph}}
 \frac{(\omega_{th}^2 - \omega^2)}{(\omega_{th}^2 - \omega^2)^2 + 
 4 \gamma^2 \omega^2}
\label{4}
\end{equation}
where $\beta = g_{ep}^2 \int g({\bf q}) d^2 q$ and ${\omega_{th}}^2 = 
(\omega_{ph} + \Delta)^2 + \gamma^2$. Solving now for the dispersion, we
 find that at low energies, the dispersion is linear with
$v^*_F = v_F/(1 + \lambda^{ep}_{sc})$ and $\lambda^{ep}_{sc}= 
\pi \beta/(\omega_{ph} \omega^2_{th})$.
Near $\omega_{th}$, the
real part of the self-energy for small $\gamma$
 nearly diverges as $\omega$ approaches $\omega_{th}$, giving rise to 
an $S$-shape dispersion. We illustrate this in Fig.~\ref{fig6}.
The $S$-shape  could in principle  be eliminated if $\gamma$ is
 very large, i.e,  the bosonic mode is almost overdamped. 
However, one can easily make sure that to avoid the $S-$shape, one requires
\begin{equation}
\frac{\gamma^2}{\omega^2_{th}} >  \frac{1}{2}~\lambda^{ph}_{sc}.
\end{equation}
 This can be relatively easily
 achieved at weak coupling, but for $\lambda^{ep}_{sc} \geq 1$, which is 
 required
 to fit the low energy renormalizaton, the $S-$shape is eliminated 
 only at nonphysical 
$\gamma >> \omega_{th}$. Furthermore, 
 for $\lambda^{ph}_{sc} >2$, the $S-$shape cannot be eliminated for any $\gamma$.
If we do not neglect the Fermi velocity at 
${\bf k}_F + {\bf q}_{ph}$, then at the lowest frequencies we obtain 
 qualitatively the same result as 
(\ref{s4}), i.e, a nearly one-sided square-root singularity. Again, it is very difficult to get 
rid of the $S-$shape form of the dispersion at strong coupling 
without requiring that the bosonic mode is totally overdamped.
  It is therefore very unlikely that the 
 interaction with an optical phonon can simultaneously account for a 
 strong Fermi velocity renormalization and give rise 
 to a non $S-$shape form of the dispersion.

\section{Conclusions and experimental comparisons}

We conclude therefore that the spin fluctuation
 scenario more likely explains the observed features in the  
electron dispersion than the phonon scenario.  Within the spin fluctuation 
scenario: (i) in the normal state, there is a crossover from a linear to a more steep 
dispersion at around $\omega_{sf}$, (ii)  below $T_c$ the
antinodal dispersion develops an $S-$shape form due to interaction with the spin resonance
with a characteristic energy of $\Delta + \omega_{res}$, (iii) the nodal dispersion below $T_c$ develops 
a kink at $2\Delta$, but there is no $S-$shape dispersion as there 
is no spin resonance  for momenta which connect a nodal point with other 
points on the Fermi surface.  
 
\subsection{a comparison with experiments}

Qualitatively, our picture of an $S-$shape dispersion in the antinodal region below
$T_c$, and a kink dispersion 
in the nodal region which is similar above and below $T_c$,
 agrees well with the data. Quantitatively, 
 $\omega_{sf} ({\bf q_0})$ and $\omega_{sf} ({\bf q_1})$  relevant for the antinodal and nodal 
dispersion, respectively, were 
 estimated to be $\omega_{sf} ({\bf q_0}) \sim 20 meV$ and 
$\omega_{sf} ({\bf q_1}) \sim 40-50 meV$ in $Bi2212$ \cite{rob}. 
Experimentally, this scale has been detected for the nodal dispersion and 
is around $50 meV$. 
The resonance frequency $\omega_{res}$ and the gap $\Delta$ in optimally
 doped $Bi2212$ are both near 
$40 meV$ \cite{pdh}, hence
 the termination of the $S-$shape in the antinodal dispersion and the 
 kink in the  nodal dispersion both occur near $80 meV$. 
Experimentally, this scale is $80 meV$ along the antinodal direction
 and $50-70 meV$ along the nodal direction \cite{sato}.  Both of these gap-related 
scales are smaller in LSCO, 
but for that material, there are no noticeable differences between the normal and 
superconducting state dispersions. This  
 implies that the effect of the superconductivity on the dispersion is very small (though
 it should be noted that most of the ``normal state" data for LSCO were actually taken
 in the pseudogap state). We note in passing that the same smallness was cited as a reason for 
 the non-observation of the resonance peak in LSCO \cite{no_res_la}.

\subsection{the doping dependence}

Finally, we discuss the doping dependence of the dispersion. 
There are few systematic studies of the doping dependence along the antinodal 
direction \cite{jc99,kim,gromko}. Our study shows that the magnitude of the $S-$shape  dispersion 
should increase with underdoping as $\chi_s ({\bf q_0})$ increases, and 
 the  coupling constant gets larger.   
Along the nodal direction,  the low energy
Fermi velocity $v^*_F \sim 1.6 eV\AA$ is relatively doping independent 
 \cite{zhou}. 
In constrast, the slope of the high energy dispersion monotonically 
increases with underdoping \cite{zhou}.
Within our theory, the low energy Fermi velocity is given by $v^*_F = 
v_F/(1+ \lambda ({\bf q_1}))$, and the coupling constant $\lambda ({\bf q_1})$
 depends on the spin-fermion coupling $g_s$ and $\chi_s ({\bf q_1})$. 
 The coupling constant is weakly doping dependent. The susceptibility does 
depend on doping via the magnetic correlation length $\xi$, but this
 dependence is non-singular for the node-antinode ${\bf q_1} \neq {\bf q_0}$. 
Once $|{\bf q_1} - {\bf q_0}| > \xi^{-1}$, the doping dependence disappears, 
and the nodal coupling $\lambda ({\bf q_1})$ saturates at a
 fixed value. Previous studies by both us and others yielded 
 $\lambda ({\bf q_1}) \sim 1$. This yields a bare velocity  $v_F \sim 3 eV \AA$ consistent with band theory. 

We believe that the increase of the high energy slope with underdoping is a separate 
effect associated with the fact that
 at high energies, the system progressively develops SDW precursors. 
Indeed, at high energies,  
 the diagonal scattering by the resonance mode at ${\bf q_0}$ cannot be neglected. 
 The argument is simple --  at high energies,
 the Green's function of an intermediate fermion can be pulled out from the 
momentum and frequency integral in Eq.~(\ref{s1}), and the 
 fermionic self-energy acquires the same functional form
 as in the SDW ordered state:
\begin{equation}
\Sigma_{SDW} ({\bf k}, \omega) \approx \frac{\Delta^2_{SDW}}{\omega - \epsilon_{{\bf k}+{\bf q_0}}}
\label{7}
\end{equation}
where $\Delta_{SDW} \propto g_s^2\int d^2 q d \Omega \chi_s ({\bf q}, \Omega)$ increases with underdoping. 
This form is valid for $|\omega| \gg \omega_0$.   
Substituting this self-energy into the dispersion relation, we 
find after simple algebra that
 for the negative energies probed in ARPES measurements, 
the maximum of the MDC dispersion   
 shifts from $\omega \approx \epsilon_{\bf k}$ to 
$\omega \approx \epsilon_{\bf k} - \Delta^2_{SDW}/(\epsilon_{{\bf k}+{\bf q_0}} - \epsilon_{{\bf k}})$.
As $\epsilon_{{\bf k}} <0$ and $\epsilon_{{\bf k}+{\bf q_0}} >0$, the correction to the velocity 
is positive, i.e., the diagonal scattering enhances
the value of the velocity  compared to $ v_F$. As $\Delta_{SDW}$ increases, this effect 
 gives rise to a progressively steeper dispersion, in agreement with the data. 

A simple way to appreciate the doping dependence of the high energy dispersion is to
realize that the MDC dispersion for a gapped state is almost vertical for 
$|\omega| < \Delta_{SDW}$, with some weak dispersion due to
coherence factors \cite{mike01}.  Thus, as the doping is decreased, and a Mott-Hubbard
pseudogap begins to develop, the high energy dispersion is expected to become
increasingly steep.

\acknowledgments

We acknowledge useful discussions with Girsh Blumberg, Juan Carlos Campuzano, Erica Carlson,
 Dan Dessau, Hong Ding, Matthias Eschrig, Peter Johnson, Adam Kaminski, and Filip Ronning.
A.C. is supported by the NSF-DMR 0240238 and M.N. by the U. S. Dept. of Energy, Office of Science,
under Contract No. W-31-109-ENG-38.  A.C. is thankful to Argonne Natl. Lab for hospitality during the
initial stages of this work.


\begin{thebibliography}{99}

\bibitem{valla}
T. Valla, A.V. Fedorov, P.D. Johnson, B.O. Wells,
S.L. Hulbert, Q. Li, G.D. Gu, and N. Koshizuka, 
Science {\bf 285}, 2110 (1999).

\bibitem{bogdanov}
P.V. Bogdanov, A. Lanzara, S.A. Kellar, X.J. Zhou, E.D. Lu, W.J. Zheng,
G. Gu, J.-I. Shimoyama, K. Kishio, H. Ikeda, R. Yoshizaki,
Z. Hussain, and Z. X. Shen, Phys. Rev. Lett. {\bf 85}, 2581 (2000).

\bibitem{kaminski}
A. Kaminski, M. Randeria, J.C. Campuzano, M.R. Norman, H.Fretwell, J. Mesot, 
T. Sato, T. Takahashi, and K. Kadowaki, Phys. Rev. Lett. {\bf 86}, 1070 (2001).

\bibitem{lanzara}
A. Lanzara, P.V. Bogdanov, X.J. Zhou, S.A. Kellar, D.L. Feng,
E.D. Lu, T. Yoshida, H. Eisaki, A. Fujimori, K. Kishio, J.-I. Shimoyama,
T. Noda, S. Uchida, Z. Hussain, and Z.-X. Shen,
Nature (London) {\bf 412}, 510 (2001).

\bibitem{johnson}
P.D. Johnson, T. Valla, A.V. Fedorov, Z. Yusof, B.O. Wells, Q. Li,
A.R. Moodenbaugh, G.D. Gu, N. Koshizuka, C. Kendziora, Sha Jian,
and D.G. Hinks, Phys. Rev. Lett. {\bf 87}, 177007 (2001).

\bibitem{ronning}
F. Ronning, T. Sasagawa, Y. Kohsaka, K. M. Shen, A. Damascelli, C. Kim, T. Yoshida,
N. P. Armitage, D. H. Lu, D. L. Feng, L. L. Miller, H. Takagi and Z.-X. Shen,
Phys. Rev. B {\bf 67}, 165101 (2003).

\bibitem{zhou}
X. J. Zhou {\it et al.}, Nature (London) {\bf 423}, 398 (2003).

\bibitem{sato}
T. Sato, H. Matsui, T. Takahashi, H. Ding, H.-B. Yang, S.-C. Wang, T. Fujii, T. Watanabe,
A. Matsuda, T. Terashima and K. Kadowaki, Phys. Rev. Lett. {\bf 91}, 157003 (2003).

\bibitem{kim}
T. K. Kim, A. A. Kordyuk, S. V. Borisenko, A. Koitzsch, M. Knupfer, H. Berger, and
J. Fink, Phys. Rev. Lett. {\bf 91}, 167002 (2003).

\bibitem{gromko}
A. D. Gromko, A. V. Federov, Y.-D. Chuang, J. D. Koralek, Y. Aiura, Y. Yamaguchi,
K. Oka, Y. Ando, and D. S. Dessau, Phys. Rev. B {\bf 68} 174520 (2003).

\bibitem{mike01}
M.R. Norman, M. Eschrig, A. Kaminski, and J.C. Campuzano,
Phys. Rev. B {\bf 64}, 184508 (2001).

\bibitem{acs}
Ar. Abanov, A. V. Chubukov, and J. Schmalian, Adv. Phys. {\bf 52}, 119 (2003).

\bibitem{ding}
M.R. Norman and H. Ding, Phys. Rev. B {\bf 57}, 11089 (1998).

\bibitem{ac}
Ar. Abanov and A. V. Chubukov, Phys. Rev. Lett. {\bf 83}, 1652 (1999)
and Phys. Rev. B {\bf 61}, 9241 (2000).

\bibitem{eschrig}
M. Eschrig and M.R. Norman, Phys. Rev. Lett. {\bf 85}, 3261 (2000);
Phys. Rev. B {\bf 67}, 144503 (2003).

\bibitem{pdh}
M.R. Norman, H. Ding, J.C. Campuzano, T. Takeuchi, M. Randeria, T. Yokoya,
T. Takahashi, T. Mochiku, and K. Kadowaki, Phys. Rev. Lett. {\bf 79}, 3506 (1997).

\bibitem{keimer}
H. F. Fong, P. Bourges, Y. Sidis, L. P. Regnault, J. Bossy, A. Ivanov,
D. L. Milius, I. A. Aksay, B. Keimer,
Phys. Rev. B {\bf 61}, 14773 (2000);
Pengcheng Dai, H. A. Mook, R. D. Hunt, and F. Do{\v g}an,
Phys. Rev. B {\bf 63}, 054525 (2001).

\bibitem{mcqueeny}
R.J. McQueeney, Y. Petrov, T. Egami, M. Yethiraj, G. Shirane, and
Y. Endoh, Phys. Rev. Lett. {\bf 82}, 628 (1999).

\bibitem{acs2}
Ar. Abanov, A. V. Chubukov, and J. Schmalian, J. Elec. Spec. {\bf 117}, 129 (2001);
A. V. Chubukov, D. Pines, and J. Schmalian, {\it The Physics of 
Superconductors} Vol 1 ed K. H. Bennemann and J. B. Ketterson (Springer,  Berlin, 2003) p 495
(cond-mat/0201140).
 
\bibitem{pines}
see e.g.,   A. J. Millis, H. Monien, and D. Pines, Phys. Rev. B \textbf{42}, 167 (1990).

\bibitem{rob}
R. Haslinger, Ar. Abanov, and A. Chubukov, Europhys. Lett. {\bf 58}, 271 (2002).

\bibitem{res_peak_theory}
M. R. Norman, Phys. Rev. B {\bf 61}, 14751 (2000) and references therein.

\bibitem{bourges}
P. Bourges, Y. Sidis, H. F. Fong, L. P. Regnault, J. Bossy, A. Ivanov and 
B. Keimer, Science {\bf 288}, 1234 (2000).

\bibitem{carbotte} see e.g., J. P. Carbotte, Rev. Mod. Phys. {\bf 62}, 1027 (1990).

\bibitem{no_res_la}
E. Demler, S. Sachdev and Y. Zhang, Phys. Rev. Lett {\bf 87}, 067202 (2001).

\bibitem{jc99}
J.C. Campuzano, H. Ding, M.R. Norman, H.M. Fretwell, M. Randeria, A. Kaminski,
J. Mesot, T. Takeuchi,
T. Sato, T. Yokoya, T. Takahashi, T. Mochiku, K. Kadowaki, P. Guptasarma,
D. G. Hinks, Z. Konstantinovic, Z.Z. Li, and H. Raffy,
Phys. Rev. Lett. {\bf 83}, 3709 (1999).

\end{thebibliography}
\end{document}